# Degradation mode estimation using reconstructed open circuit voltage curves from multi-year home storage field data

Jan Figgener[*,a,b,c,d,e], Jakob Bors[b,c,d,e], Matthias Kuipers[b,c,d,e], Felix Hildenbrand[a,b,c,d], Mark Junker[a,b,c,d], Lucas Koltermann[a,b,c,d], Philipp Woerner[b,c,d], Marc Mennekes[b,c,d,e], David Haberschusz[a,b,c,d,e], Kai-Philipp Kairies[b,c,d,e], Dirk Uwe Sauer[a,b,c,d,e,f]

[a] *Center for Ageing, Reliability and Lifetime Prediction of Electrochemical and Power Electronic (CARL), RWTH Aachen University, Germany*

[b] *Institute for Power Electronics and Electrical Drives (ISEA), RWTH Aachen University, Germany*

[c] *Institute for Power Generation and Storage Systems (PGS), E.ON ERC, RWTH Aachen University, Germany*

[d] *Juelich Aachen Research Alliance, JARA-Energy, Germany*

[e] *ACCURE Battery Intelligence GmbH, Germany*

[f] *Helmholtz Institute Münster (HI MS), IMD 4, Forschungszentrum Jülich, Germany*

[*] *Corresponding author at Institute for Power Electronics and Electrical Drives (ISEA)*

*Mathieustraße 10, Aachen, Germany | Mail: jan.figgener@isea.rwth-aachen.de*

## Abstract

A battery's open circuit voltage (OCV) curve can be seen as its electrochemical signature. Its shape and age-related shift provide information on aging processes and material composition on both electrodes. However, most OCV analyses have to be conducted in laboratories or specified field tests to ensure suitable data quality. Here, we present a method that reconstructs the OCV curve continuously over the lifetime of a battery using the operational data of home storage field measurements over eight years. We show that low-dynamic operational phases, such as the overnight household supply with electricity, are suitable for recreating quasi OCV curves. We apply incremental capacity analysis and differential voltage analysis and show that known features of interest from laboratory measurements can be tracked to determine degradation modes in field operation. The dominant degradation mode observed for the home storage systems under evaluation is the loss of lithium inventory, while the loss of active material might be present in some cases. We apply the method to lithium nickel manganese cobalt oxide (NMC), a blend of lithium manganese oxide (LMO) and NMC, and lithium iron phosphate (LFP) batteries. Field capacity tests validate the method.

## 1 Introduction

This paper builds upon our previously presented capacity estimation method in Figgener et al. [1] and its corresponding field dataset comprising 21 home storage systems over up to eight years [2]. With this substantial extension, we publish a method for estimating open circuit voltage (OCV) curves and discuss the observed capacity loss by degradation mode investigation using incremental capacity analysis (ICA) and differential voltage analysis (DVA).

**OCV curve.** A battery's OCV curve describes the relation between state of charge (SOC) and cell voltage at an open circuit and can be seen as its electrochemical signature. It enables deducing the SOC from the OCV and vice versa and is therefore often used as a simple method to estimate SOC [3–5]. The OCV curve is a characteristic of the active material in both anode and cathode, cell balancing, and temperature [6,7]. As active material and balancing changes over battery lifetime, the OCV curve shifts. Tracking the OCV curve is a widespread approach to gaining insights into degradation modes [8,9].

**OCV measurements.** Measuring OCV requires the battery to be in a relaxed state, excluding all influence of kinetic overvoltage on measured battery voltage. Hence, it cannot be measured directly during operation. Two standard techniques to determine OCV curves are low-current and incremental OCV measurements [10]. With a focus on investigating changes in OCV curves, most researchers use low-current OCV tests, where cell voltage is measured at low current rates (C-rates) down to C/20 [11–16]. Both approaches require high measuring precision and controlled temperature conditions and, thus, are usually conducted under laboratory conditions. The resulting curves are called quasi-OCV (qOCV) curves. Other publications use higher C-rates, some of which intend to mimic real-life scenarios, at C-rates between C/3 and 1C [17–24]. The impact of C-rates on these measurements is investigated in [10] and [25]. Instead of monitoring battery aging by repeating these measurements, the development of OCV curves for artificial scenarios can be simulated using models based on initial open circuit potential measurements of the electrodes, as shown in [8] and [26].

**OCV analysis.** OCV curve tracking to analyze field data would improve state of health (SOH) estimation and prediction of battery systems by using knowledge gained in laboratory tests. Only a few publications exist to this date discussing the creation of OCV curves directly from monitoring field operation. In [27], EV driving voltage data is transformed into an OCV curve by selecting voltage measurements at low current rates (< C/20) and mapping them to an SOC. The resulting curve, however, does not show the level of detail required for in-depth aging analyses and is only used for battery modeling. EV charging data can also be used to analyze the OCV [28,29] and is suitable as charging conditions are similar. Other publications present concepts to create OCV curves from field data using load profiles in laboratory tests [30]. Some use partial charging data at varying C-rates representing a realistic charging process to reconstruct the OCV curve using a battery model [31,32] or neural network [33]. In [34,35], the OCV curve is reconstructed by determining the length of individual voltage plateaus in the curve from normal charging or discharging processes. One of the main challenges to overcome when recreating OCV curves from field data is the high noise level in these measurements, which can be adjusted using Gaussian filtering [15]. The impact of noise and different approaches to reduce its influence on these measurements are presented in [36]. Once the battery OCV curves are available over the lifetime of an aging battery, their development can be analyzed directly or by ICA or DVA.

**Incremental capacity analysis.** The ICA is a method to investigate the derivation of the OCV curve, which contains information about the electrochemical properties of the materials and can be analyzed during aging to obtain information about their changes [17,37]. By differentiating the capacity towards the voltage, voltage plateaus of the OCV curve are translated into clearly identifiable peaks in the ICA curve [17,37]. The development of the intensity and position of these peaks contain information regarding the battery's aging and the underlying degradation modes [18,37,38]. The ICA curve is determined as the deviation of the charge concerning the voltage [15,37]. Since the OCV curve is often only available as discrete values, the ICA is usually computed as the gradient using Equation (1-1) [39].

$$\frac{dQ}{dV} \approx \frac{\Delta Q}{\Delta V} \qquad (1\text{-}1)$$

**Differential voltage analysis.** The DVA can be used to investigate the utilization and balancing between the electrodes [40]. Here, the OCV curve is differentiated concerning SOC using the gradient function in Equation (1-2) [39]. Peaks in the DVA curve represent transitions between two voltage plateaus in the OCV, and their positions can be used to investigate cathode and anode degradation separately, as well as shifts in the electrode balancing [40,41].

$$\frac{dV}{dQ} \approx \frac{\Delta V}{\Delta Q} \qquad (1\text{-}2)$$

**Degradation modes.** The three commonly investigated degradation modes (DMs) across the presented literature are loss of lithium inventory (LLI) and loss of active material of both the anode and cathode ($LAM_{NE}$ and $LAM_{PE}$, respectively) [37]. In the case of blended cathodes, the losses of the different active materials can be investigated individually [37]. In [8] (laboratory data) and [34] (EV charging data), battery models based on initial half-cell measurements are fitted to the measured OCV curves, quantifying the number of different DMs inside the battery. Many publications use ICA for aging analyses, with [21,25,28,29,36,42–46] focusing on SOH estimation and [14,16–20,26,38,47–51] using it as a method to identify DMs in varying scenarios. Publications using DVA generally investigate electrode degradation and changes in electrode balancing [11,13,41,52–55]. Since ICA and DVA allow varying insights into battery aging, they can also be combined to gather additional information about DMs, as done in [12,15,23,24,56,57].

**Features of interest.** A recent addition to these methods is investigating varying features of interest (FOIs) within the characteristic IC and DV curves and their correlations with individual DMs. In the case of the ICA, FOIs usually are the position or intensity of specific features like minima or maxima. In the case of the DVA, the distances between different characteristic peaks are usually used as FOIs. Their choice depends on the technology of the investigated battery, which determines the shape of IC and DV curves. A large study examining such correlations of several FOIs in both the IC and DV curves of different battery chemistries can be found in [37,56].

**Scientific contribution.** While many publications focus on ICA, DVA, and the way DMs can be quantified using them, these findings have not yet been applied to home storage system field data to the best of our knowledge. Two publications at the time of writing are presented by Dubarry et al. in [58] and their extended analysis [59]. In their studies, they successfully apply ICA to synthetic data of HSSs, identify clear sky conditions as suitable for the analysis of charging curves, and state that there is a lack of field data [58]. Using our previously published dataset of HSS operational data in private homes over up to eight years [1,2], this work contributes to recreating OCV curves from real-world field data. The method developed helps to gain insights into the battery aging of HSSs in field operation by applying laboratory-proven knowledge to field data analytics. It can be used by battery storage operators, battery analytic companies, and researchers to analyze battery aging using existing field data.

## 2 Methodology

The methodology of this work is displayed in Figure 1. The OCV curves from home storage systems are reconstructed from eight years of field data published with our previous publication in Figgener et al. [1,2] using several filters and voltage correction. The obtained qOCV curves are analyzed and used for both ICA and DVA to determine the occurring DMs by tracking FOIs derived from literature. The qOCV curves are validated by field capacity tests conducted in [1].

### 2.1 Dataset and measurement

The measurements have been conducted over a period of up to eight years, from 2015 to 2022 and the dataset comprises 21 home storage systems, 106 system years and around 14 billion datapoints. The measured quantities are the system-level battery voltage, current, power, and housing temperature. All quantities are measured with a sample rate of one second. [1,2]

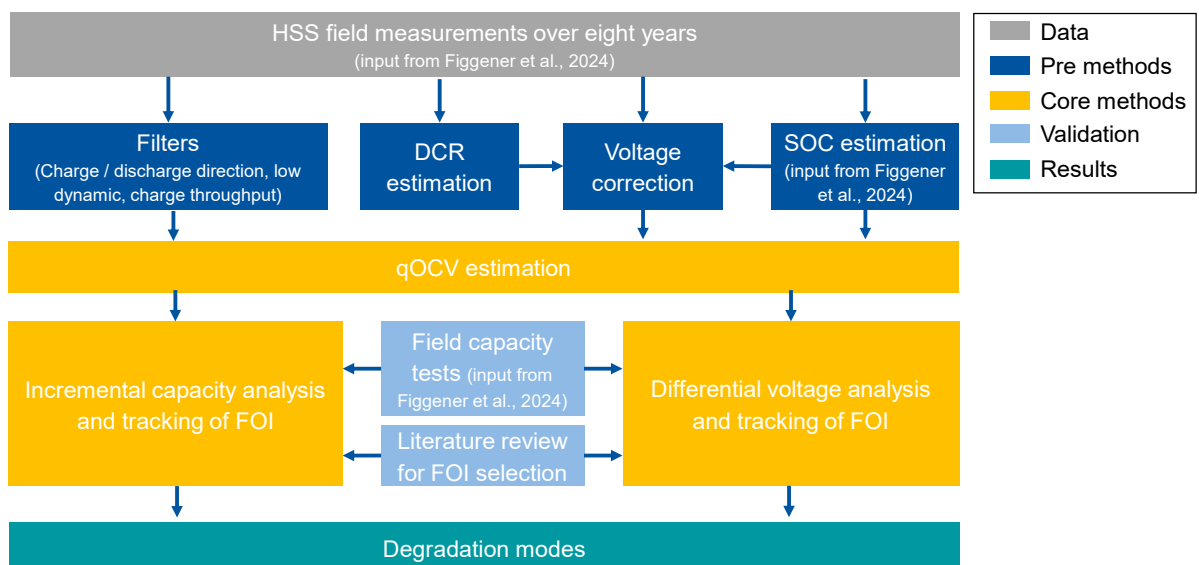

Figure 1: Methodology to estimate the qOCV curve and identify the degradation modes through the incremental capacity analysis and the differential voltage analysis.

The dataset comprises many different HSS products. The main differences are the system design in terms of energy and power ratings and the battery chemistry used. To account for both, the nomenclature is a combination of system size and chemistry, according to [1]. The smallest HSSs have a battery energy of around 2 kWh and use a blend of lithium manganese oxide (LMO) and lithium nickel manganese cobalt (NMC). These systems are referred to as "Small$_{LMO}$" HSSs, as the average home storage system was about 9 kWh in 2022 [60]. There are other systems in the range of around 8 kWh to 13 kWh. These systems are divided into HSSs with different NMC and different lithium iron phosphate (LFP) batteries, referred to as "Medium$_{NMC}$" and "Medium$_{LFP}$" HSSs, respectively.

## 2.2 qOCV curve estimation

The algorithm used in this work to derive qOCV curves from field data is based on detecting low dynamic dis-/charging phases to have as little overvoltage variation as possible. An additional overvoltage correction is developed by estimating the direct current resistance (DCR). Suitable operational phases can be combined to form a complete qOCV curve. In the following, the individual steps of the algorithm will be presented in detail.

### 2.2.1 Direct current resistance estimation

The DCR is needed for overvoltage correction. It is calculated according to Equation (2-1) with $V_1$ and $I_1$ being the voltage and current values at the beginning of the pulse, and $V_2$ and $I_2$ at its end. This calculation is similar to EV battery analyses done in [61]. As the DCR is dependent on temperature and SOC, both values are considered. The C-rate dependency is neglected because the applied dynamics filter automatically accounts for similar C-rates (see Chapter 2.2.2).

$$DCR(SOC, T) = \frac{V_2 - V_1}{I_2 - I_1} = \frac{\Delta V}{\Delta I} \quad (2\text{-}1)$$

$$\text{for all values of SOC and temperature (T)}$$

Figure 2 presents the resulting DCR estimation for an exemplary system. It showcases how the DCR tends to increase at low SOCs and low temperatures, which aligns with findings in the literature [62].

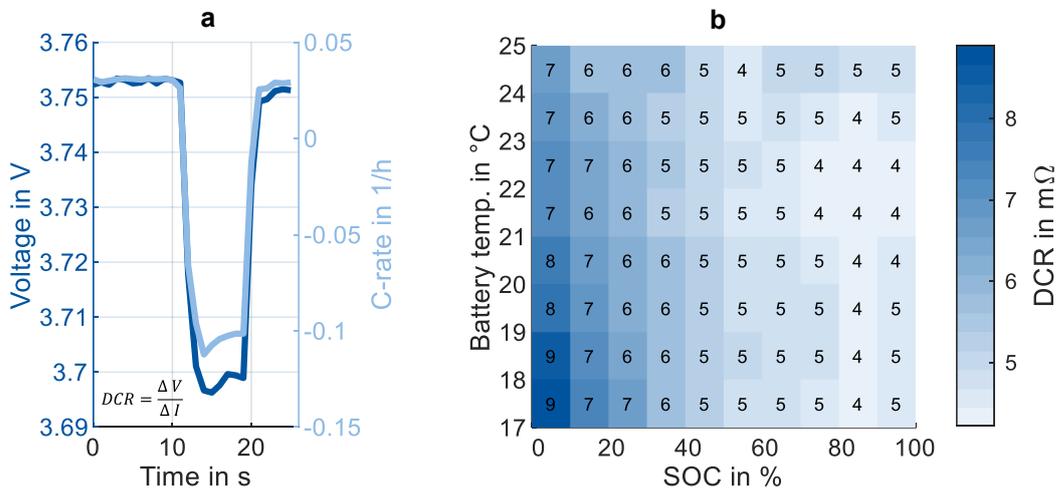

Figure 2: **a)** Identified current pulse and voltage response to calculate the DCR by dividing the voltage delta by the current delta. **b)** Estimated DCR values for an exemplary Small$_{LMO}$ system (cell level) according to SOC and temperature.

Throughout most operating conditions, the DCR values of the shown system remain around 5 mΩ, but they can reach up to 9 mΩ at low temperatures and low SOC levels. The DCR and its dependency on SOC change over time, while the temperature influence stays approximately similar [62]. Thus, the DCR look-up tables must be recalculated regularly. Appendix, Figure 6, shows the long-term trend for one exemplary system. To give a simplified overview, Appendix, Figure 7, depicts the linear gradients of yearly DCR values for different SOC ranges in a defined temperature range. The DCR increases are higher at the SOC ranges and there are substantial differences between the three system types. While the mean increase for Small$_{LMO}$ systems is 24.6 percentage points per year (pp/a) for an SOC below 10 %, the mean values are around 15 pp/a for the SOC ranges of 40 % to 60 % and above 90 %. The Medium$_{NMC}$ HSSs show increased aging both for low (mean 5.2 pp/a) and high SOCs (mean 4.4 pp/a), while the aging is less for medium SOCs (1.7 pp/a). The Medium$_{LFP}$ HSSs show only small mean DCR increases from 0.8 pp/a to 1.8 pp/a with higher gradients for higher SOCs.

### 2.2.2 Operational phase detection

**Charge / discharge filter.** The qOCV curves are created both for charge and discharge direction. A charge phase is identified by finding periods during which the current sign is strictly positive. In contrast, the discharge phase current needs to be strictly negative. This filter accounts for overvoltage direction and hysteresis.

**Charge throughput filter.** The charge throughput filter reduces the charge and discharge phases to those containing voltage information over a predefined minimum SOC range. The SOC range covered by an individual phase is determined by the absolute difference in SOC from the beginning of the phase to the end of the phase. The limit is set to a percentage value of 5 % of the battery capacity.

**Dynamics filter.** A dynamics filter is implemented to avoid ripples in the qOCV curve caused by high-dynamic operation. The dynamic is defined as the absolute short-term change in current, which can be directly determined from the measurement data. By setting an upper limit of 10 % of the C-rate to this dynamic, the data can be reduced to periods consisting of low-dynamic data only. Figure 3a shows the identified voltage phases at low-dynamic C-rates overnight.

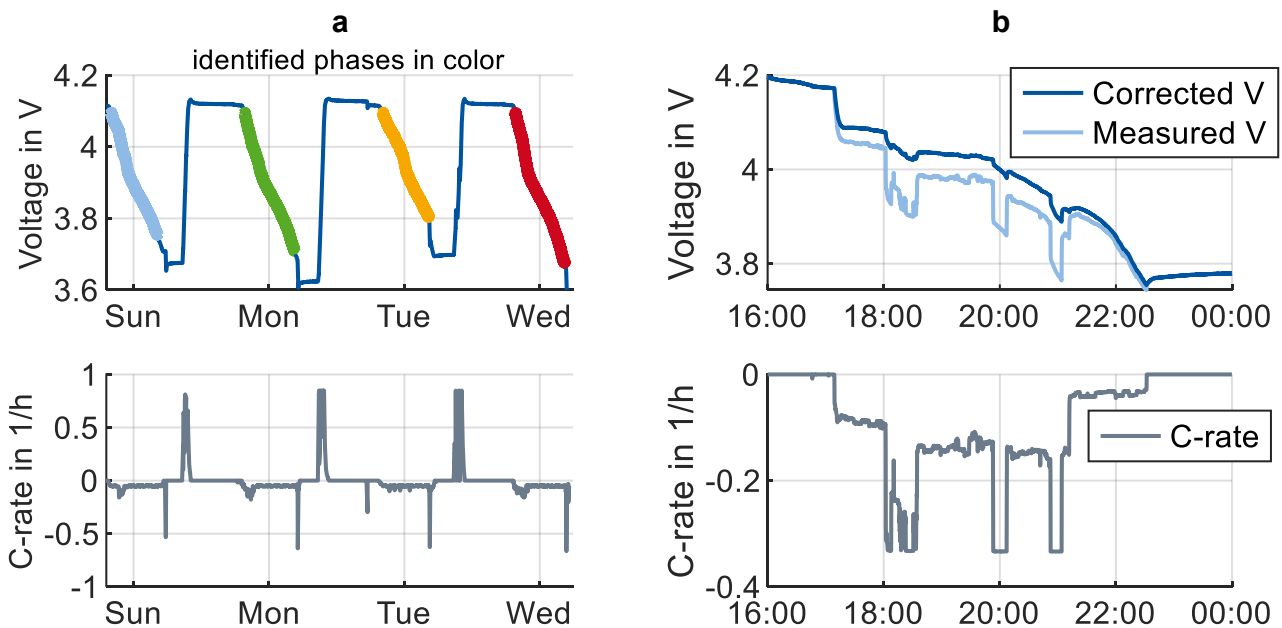

Figure 3: **a)** Identified low-dynamic current phases represent partial qOCV curves. **b)** Exemplary overvoltage correction. The corrected voltage is quite smooth, especially for low-dynamic phases.

**Overvoltage correction.** Figure 3b shows an exemplary overvoltage correction according to Equation (2-2). The voltage correction is not perfect, but it leads to smoother voltage courses, especially in low-dynamic operational phases.

$$V_{\text{corrected}}(t, \text{SOC}, T) = V_{bat}(t) - I_{bat}(t) \cdot \text{DCR}(\text{SOC}, T) \qquad (2\text{-}2)$$

### 2.2.3 Combination of partial qOCV curves

A detected low-dynamic phase represents a partial qOCV curve on its own if the corrected voltage is plotted against the corresponding SOC. Combining many of these partial qOCV curves leads to an overall qOCV curve for the whole cell, which can be updated continuously. Figure 4a visualizes this process for a limited number of exemplary discharge phases. Sometimes, SOC errors occur, leading to a horizontal deferral of single identified phases. In this case, the SOC values are shifted horizontally to the mean value of all partial qOCV curves. This way, the voltage course is maintained as only the SOC values are adjusted.

The resulting qOCV curves from individual operational phases cover different SOC and voltage ranges. The overall qOCV curve is computed by taking the mean SOC of all available phases at small voltage steps, resulting in a high-resolution qOCV curve (see Figure 4b). The qOCV curves are created separately from the discharging or charging phases to account for remaining overvoltage and hysteresis effects. The actual OCV curve of the battery lies between the charge and discharge qOCV curves.

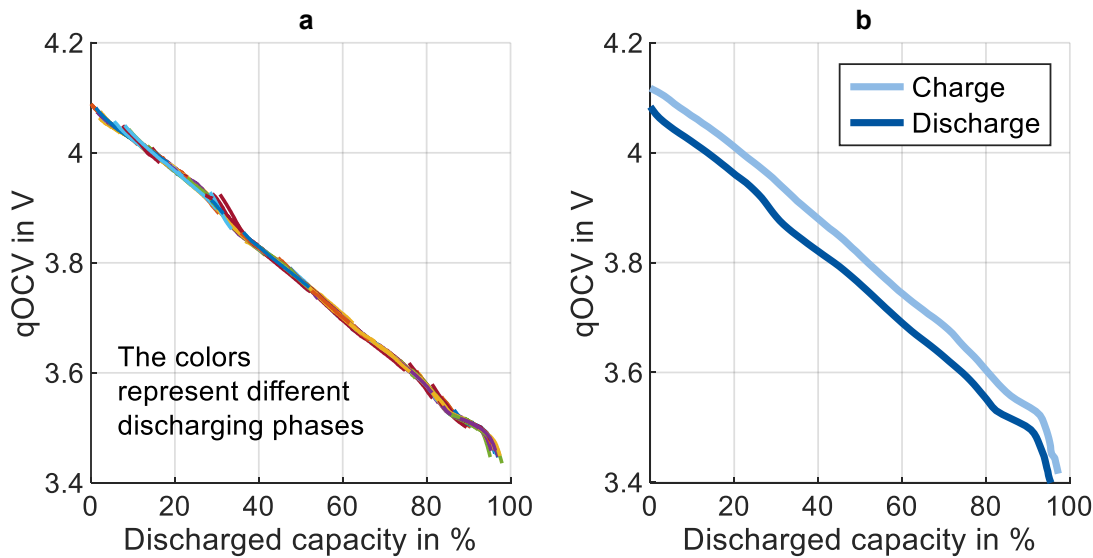

Figure 4: **a)** Combination of partial qOCV curves from low-dynamic discharge current phases. Their mean is taken as the estimated qOCV curve of a specific period. **b)** Resulting qOCV curves for charge and discharge. The discharge curves have higher quality.

The discharge phases show lower dynamics and lower C-rates (see Figure 3a and [1]) than the charge phases, which lead to higher-quality qOCV curves. Thus, only the discharging qOCV curves are used from here on. However, acceptable results can be obtained from the charge phases as well.

The investigations performed on the qOCV curves are focused on analyzing aging behavior. For this reason, the data is split into periods so that independent continuous qOCV curves are created for each of them. Depending on the data quality and operation of the system, a different minimum period is required to acquire enough phases to create a qOCV curve. To illustrate the methodology, the continuous qOCV curves in this work are created yearly. Note that this interval could be shorter, and sometimes, a month is enough to recreate a qOCV curve using this methodology if the HSS is cycled regularly.

### 2.3 Application of ICA and DVA

**IC and DV curves.** To analyze the qOCV curves' characteristics, both the incremental capacity analysis (ICA) and differential voltage analysis (DVA) are applied. The IC and DV curves are obtained by numerically differentiating the continuous qOCV curve concerning the voltage and SOC using Equation (1-1) and (1-2). Before determining the derivative, the qOCV curve is smoothed using a Gaussian filter, which has been proven to be a good method according to literature [15,37]. This smoothing is required since the investigated features, such as voltage plateaus in the qOCV curve, are quite sensitive to noise and trembling [15,37].

**FOI tracking.** To determine the contribution of different DMs to battery aging, FOIs are selected from literature, for which a correlation between their development and specific DMs could be determined. These are tracked automatically by identifying the corresponding extrema in specified voltage ranges. Since different chemistries show different characteristic features in their IC and DV curves, FOIs are chosen individually for each technology. While the selected ICA FOIs are mainly based on Dubarry and Anseán [37] with additional information from references [15,19,20,24,26,38,56,63], the DVA FOIs are selected from various literature [13,15,24,41,55,63–67].

**Correlation of FOI and SOH.** The tracked change in FOI for ICA (intensity and position), and for DVA (distance) is correlated with the capacity decrease obtained from Figgener et al. [1] to investigate a possible relationship between capacity fade and FOI shift. In this context, the so-called "p-values" are calculated. In correlation analysis, the p-value is a statistical metric that helps assess the significance of the observed correlation coefficient between two variables [68]. It quantifies the probability of obtaining a similar correlation as the one observed in the data, assuming no true correlation exists. A low p-value of typically less than 0.05 indicates that the correlation is statistically significant, suggesting that the relationship between the variables is not due to random chance [68]. In contrast, a high p-value indicates that the observed correlation could have occurred by chance alone and is not statistically significant [68]. Thus, the p-values are used to make informed decisions about the strength and validity of correlations.

## 3  Results and discussion

This chapter discusses the development of the qOCV discharge curves using ICA and DVA. For conciseness, this chapter limits the method presentation to the Small$_{LMO}$ systems while the appendix presents its functionality on Medium$_{NMC}$ and Medium$_{LFP}$ systems. This focus is set, as all six Small$_{LMO}$ systems use exactly the same battery type while the medium systems consist of different products with varying cells. From here onward, the terms qOCV and OCV are used as synonyms for simplification.

### 3.1  Detailed analysis of Small$_{LMO}$ systems with LMO/NMC blend

Figure 5 shows the OCV, IC, and DV curves of an exemplary Small$_{LMO}$ system, and the FOI tracking of all six of these systems.

**OCV.** The LMO/NMC OCV curve (see Figure 5a) shows the characteristic of an approximately linear voltage increase with a steeper beginning. To illustrate the capacity fade, this chapter normalizes the state of charge by the nominal capacity, which is why OCV values of aged systems do not reach 100 % of SOC$_{Nominal}$. With ongoing aging, the curves show a horizontal shift to lower SOC values as the remaining capacity decreases. The capacity fade is quantified by the difference of the SOC from year one to year seven at the EOC charge voltage of around 4.15 V. This capacity decrease amounts to 17 % over the measurement period. The voltage at low SOC values decreases from about 3.4 V to 3.3 V. This effect is not due to aging and can be attributed to the decrease of EOD voltage controlled by the BMS, showing real-world diagnostic challenges [1]. It leads to a slight shift of the OCV curves to the right, showing the gain in capacity due to EOD voltage reduction.

**ICA.** Figure 5b shows the determined IC curves of the same Small$_{LMO}$ system over the monitored time span. FOI 1 shows a shift toward higher voltages from 3.5 V to 3.55 V and a decrease in intensity from 225 %Q/V to 150 %Q/V. The shift corresponds to capacity loss, indicating LLI [9,20,37,39]. The intensity of FOI 2 is declining over time, which can be attributed mainly to LLI, but could also include LAM$_{NE}$ and LAM$_{PE}$ [37]. FOI 3 shifts to higher voltages and slightly higher intensity, confirming the occurrence of LLI [37]. In addition, the intensity decrease of FOI 4 could indicate the presence of LAM$_{NE}$ and LAM$_{PE}$ [37]. Overall, LLI seems to be the dominant DM based on the ICA conducted.

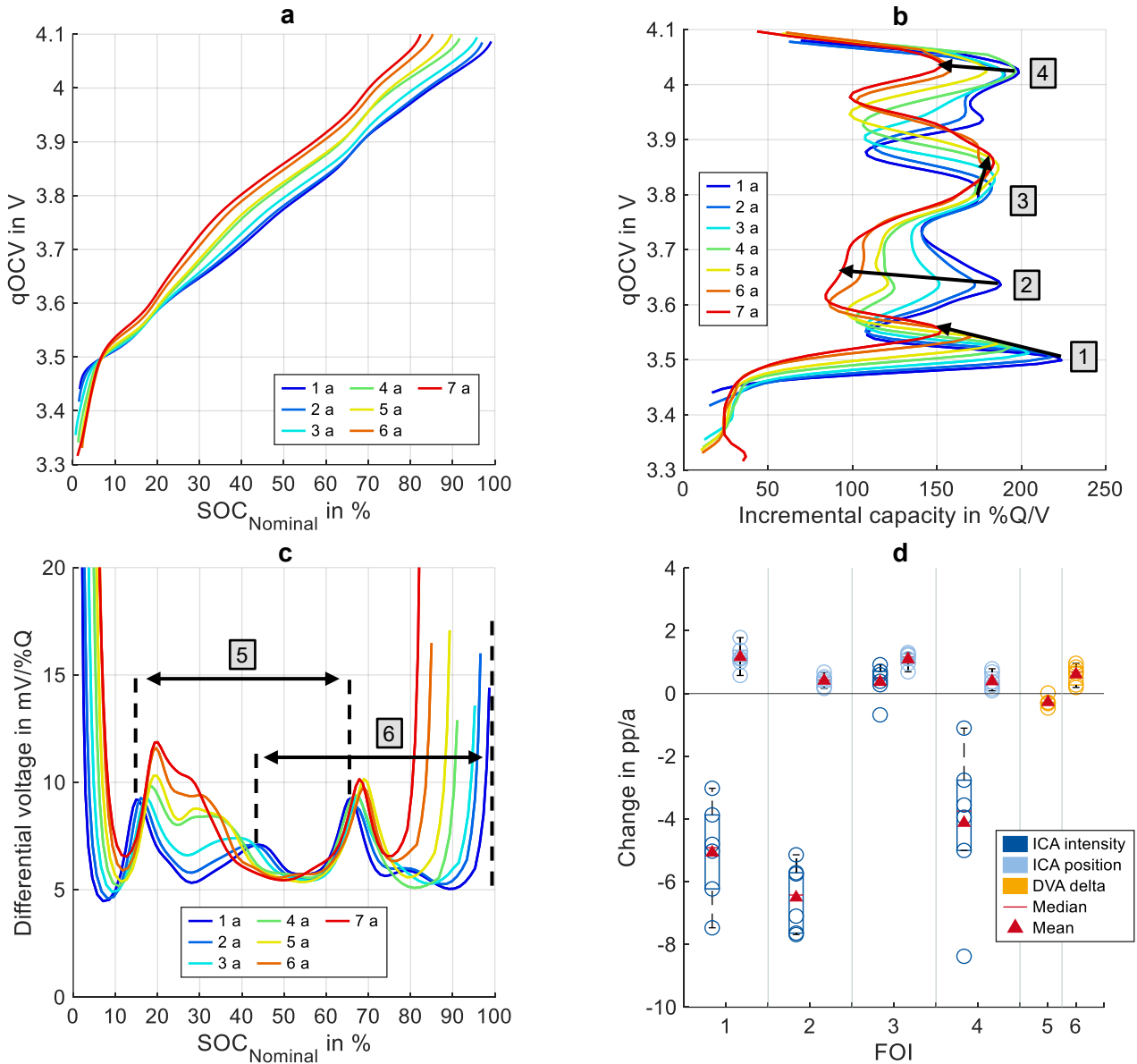

Figure 5: LMO/NMC analyses. **a)** qOCV, **b)** ICA, **c)** DVA of one exemplary Small$_{LMO}$ system, and **d)** FOI tracking of all six Small$_{LMO}$ systems. SOC values normalized by nominal capacity. FOI change is a linear fit of FOI shift over the lifetime in percentage points per year (pp/a). For ICA: Intensity normalized by highest peak and position normalized by voltage range. For DVA: Distance normalized by SOC range. Note that the Small$_{LMO}$ systems lower their EOD voltage, leading to slightly different voltage values for the same SOC. Peaks chosen from literature are clearly observable.

**DVA.** The steep rise at both ends of the DV represents the fully charged and discharged states (see Figure 5c). The LMO/NMC DV curve shows three characteristic peaks resulting from peaks in the underlying electrode DVs [63,64]. The peak starting at around 65 % to 70 % SOC represents the transition from the medium to the low voltage plateau of the anode potential [64]. The sharp peak at middle SOCs results from the cathode potential and allows conclusions about the cathode state [64]. FOI 5 is the distance from the graphite peak at around 65 % to the first peak at 10 % to 20 % SOC. It is reduced slightly, indicating a small decrease in storage capability of the anode in the form of LAM$_{NE}$ [41,56]. The cathode feature, represented by FOI 6, would indicate LAM$_{PE}$ if the distance decreased between the peak around 45 % SOC and the steep DV at the fully charged state [63]. However, the 45 % SOC peak changes in shape and becomes difficult to track. The assumed distance stays approximately constant and does not lead to clear findings. Overall, the DVA supports the assumption originating from the ICA that LLI is the dominant DM. The capacity loss of the anode does not correspond to the loss of battery capacity since the electrodes usually have excess active material, especially on the anode side [8]. Rather, the loss of capacity seen in the OCV curve can mainly be attributed to LLI.

**FOI tracking all Small$_{LMO}$ systems.** Figure 5d shows the aggregated FOI tracking for all six Small$_{LMO}$ systems. These systems show predominately LLI, while LAM$_{NE}$ and LAM$_{PE}$ could also occur in notably lower quantities. The detection of LLI is supported through three FOIs: the shift from FOI 1 and FOI 2 to lower intensities and higher voltages, and the shift of FOI 3 to higher intensities and higher voltages. Concerning the anode, small amounts of LAM$_{NE}$ may be detected through FOI 2 and the distance shrink of FOI 5. Regarding the cathode, the development of FOI 4 to a lower intensity and higher voltage could indicate LAM$_{PE}$. However, FOI 6 leads to the assumption that LAM$_{PE}$ is likely not prominent.

**Correlation of FOI and SOH.** In the following, the individual FOI shifts of all Small$_{LMO}$ systems are correlated with their respective SOH values from Figgener et al. [1]. In addition to the correlation coefficient, the corresponding p-values are shown in Table 1, representing the probability of the estimated correlation being wrong. In the case of Small$_{LMO}$ systems, all p-values except for FOI 3's position are smaller than 0.05, showing high confidence in the correlation coefficients. Using these, the highest correlation with SOH development can be seen for FOI 1 (0.91, -0.95). FOI 2 strongly correlates with SOH, especially its intensity (0.91). FOI 3's position has with correlation coefficient of -0.89. Overall, FOI 1 to FOI 3 depict a correlation of LLI with SOH. FOI 4 also suggests a correlation between LAM and SOH. FOI 5 and FOI 6 do not show a high correlation with SOH.

Table 1: Correlations of LMO FOIs with SOH development.

| FOI | 1 | | 2 | | 3 | | 4 | | 5 | 6 |
|---|---|---|---|---|---|---|---|---|---|---|
| Tracking | Int | Pos | Int | Pos | Int | Pos | Int | Pos | Dist | Dist |
| Corr. SOH | 0.91 | -0.95 | 0.91 | -0.78 | -0.03 | -0.89 | 0.85 | -0.74 | 0.34 | -0.42 |
| p-value | < 0.05 | < 0.05 | < 0.05 | < 0.05 | 0.85 | < 0.05 | < 0.05 | < 0.05 | < 0.05 | < 0.05 |

## 3.2 Summary of NMC and LFP systems

Figure 8 and Figure 9 in the appendix show that the method also works for Medium$_{NMC}$ and Medium$_{LFP}$ systems, respectively. For the Medium$_{NMC}$ system, LLI shows the strongest correlation with SOH development. For the Medium$_{LFP}$ system, the method does not reach high certainty, even though the results indicate that the occurrence of LLI is most likely. The less accurate LFP analysis is probably due to slightly inaccurate SOC and OCV estimates due to its flat voltage curve. Although the FOIs presented in literature are obtained, the method's accuracy combined with the curves' smoothing is insufficient to use the minimal shifts in FOIs for statistically meaningful identification of correlations between DMs and SOH development.

## 3.3 Validation by field capacity tests

Appendix, Figure 10, shows that the conducted field capacity tests from [1,2] can be used to validate the obtained IC and DV curves. However, ICA and DVA are quite sensitive to small noise in measurement data. Thus, the qOCV curve obtained through the presented method is considered more reliable than the field capacity tests that rely on a single measurement at the systems' highest possible C-rate.

# 4 Conclusion

This paper presents a method to identify the degradation modes responsible for capacity loss and increase in internal resistance from multi-year home storage field measurements. The developed method first reconstructs the qOCV curves from field data. For this purpose, partial low-dynamic operational phases such as clear-sky conditions during charge or steady overnight electricity supply of households during discharge are combined. Both charge and discharge phases show reasonable results, although the resulting curves from discharge phases show higher accuracy than those from the charge phases. After reconstructing the annual qOCV curves, the change of the curves over time is analyzed. Purely from the qOCV curves, it can be seen that the capacity is decreasing. The methods of incremental capacity analysis and differential voltage analysis are applied for degradation mode estimation. Here, features of interest derived from literature can be identified in the obtained curves for all three technologies studied (blend of LMO/NMC, NMC, and LFP). While the method shows good accuracy for LMO/NMC and NMC systems, the detailed analysis of LFP systems is challenging due to the flat voltage curve. The degradation mode LLI is dominant for the analyzed systems, while LAM$_{NE}$ and LAM$_{PE}$ could be present in some cases. Conducted field capacity tests serve as validation of the method.

## 5 Acknowledgment

Parts of the results were obtained within the research projects "WMEP PV-Speicher" (funding number 0325666) and "WMEP PV-Speicher 2.0 (KfW 275)" (funding number 03ET6117)", both funded by the German Federal Ministry for Economic Affairs and Climate Action (BMWK), and "Betterbat" (funding number 03XP0362B), funded by the Federal Ministry of Education and Research (BMBF).

## 6 Competing interests

As stated in the paper, the authors J.F., J.B., M.K., M.M, D.H., K.-P.K., and D.U.S. are shareholders and/or employees of ACCURE Battery Intelligence GmbH, which offers commercial battery diagnostics. The company is a spin-off of RWTH Aachen University, where the research was conducted. The other authors declare no competing interests.

## 7 Author contributions

Overall, J.F. led the research during his dissertation on home storage. J.F. and D.U.S. conceptualized the research. J.F., J.B., and M.K. developed the methodology, while F.H., M.J., L.K., and D.U.S. supported this task. J.F. and J.B. developed software to analyze and evaluate the data, while P.W., M.M., and D.H. supported this task. J.F. wrote the original paper draft and created the figures and tables with support from J.B., while M.K., F.H., M.J., L.K., D.H., K.-P.K., and D.U.S. reviewed and edited the paper. J.F. and D.U.S. supervised the research. J.F., K.-P.K., and D.U.S. did the main funding acquisition and project administration for RWTH Aachen University.

# 8  Appendix

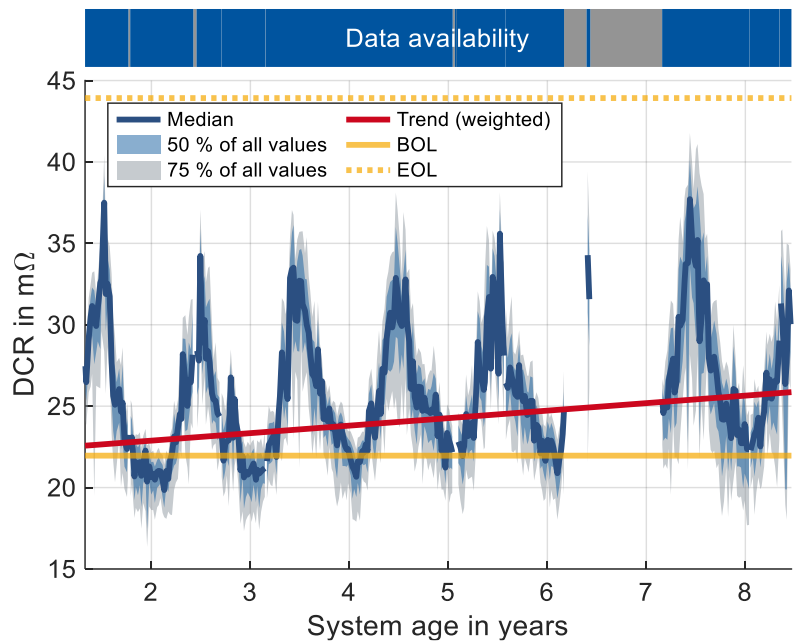

Figure 6: Development of the DCR of an exemplary Medium$_{NMC}$ (system level). The effect of different SOC and temperature values is visible. In the summer, high temperatures and elevated SOCs lead to a low DCR. In the winter, low temperatures and small SOCs lead to an increased DCR. The fit accounts for all estimates and is mainly influenced by estimates during summer due to the higher operational hours. Overall, the trend increases by less than 1 mΩ/a in the case shown, which corresponds to a relative increase of around 2.5 pp/a. Thus, the HSS does not come close to its EOL, which is defined as a doubling in internal resistance. Note that other systems do not show such a strong dependency on temperature.

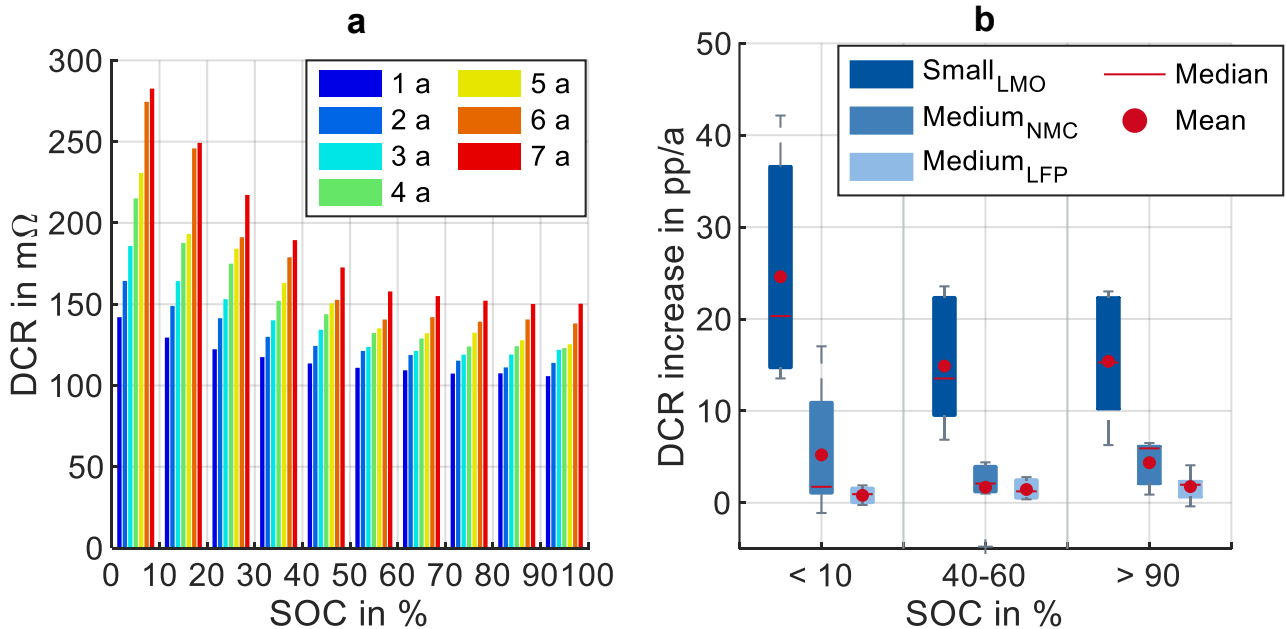

Figure 7: **a)** SOC-dependent DCR development for exemplary an exemplary Small$_{LMO}$ system (system level and mean values) plus the average annual increase in percentage points. Temperature range is 20 °C to 25 °C. The legend contains information about the system age. **b)** Linear gradients of DCR increase to simplify the influence of battery aging. The Small$_{LMO}$ systems show significantly larger DCR increases than the Medium$_{NMC}$ and Medium$_{LFP}$ systems.
11

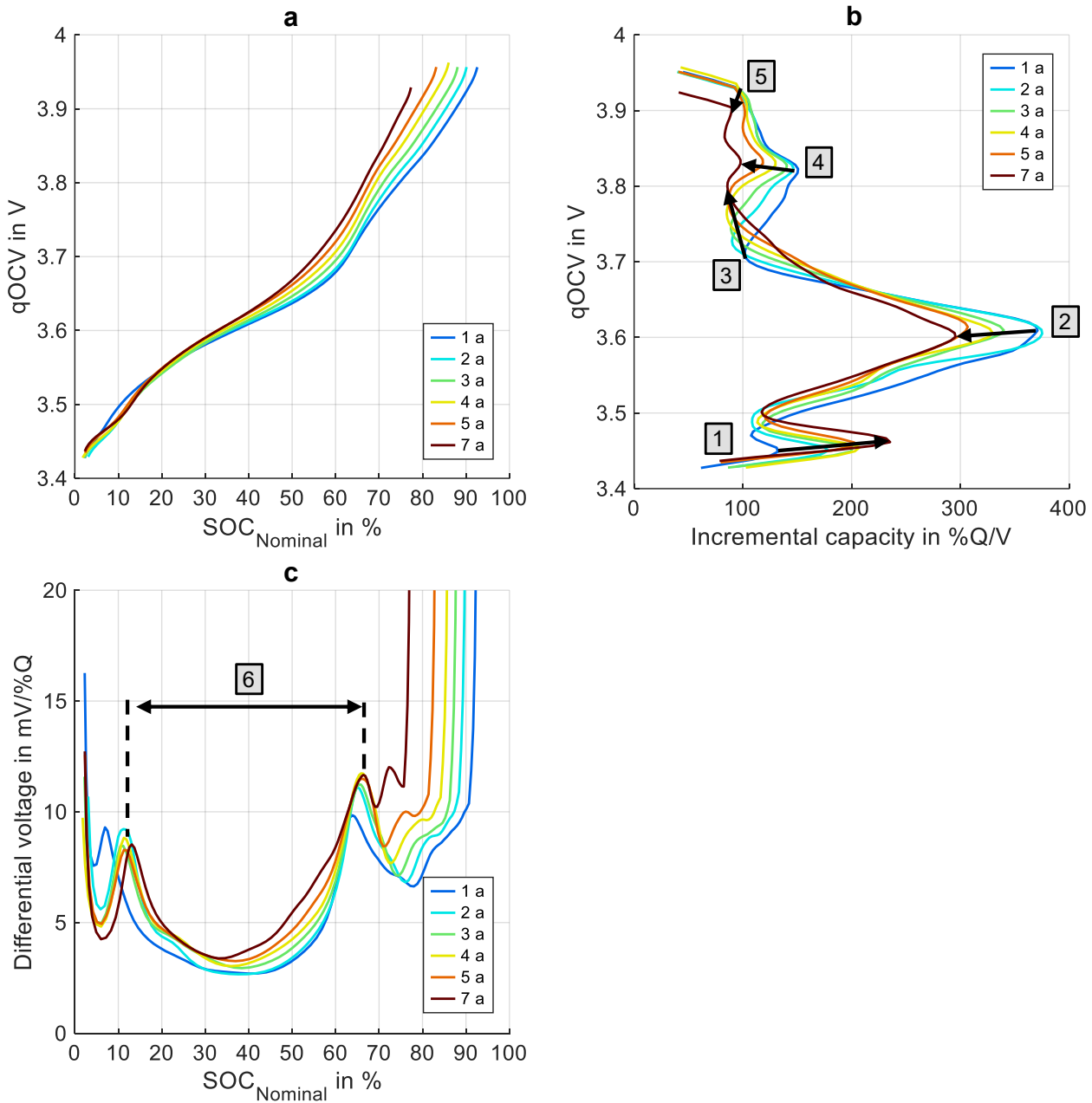

Figure 8: NMC OCV analysis of an exemplary Medium$_{NMC}$ system. Year six is missing due to measurement problems. SOC values normalized by nominal capacity.

**a) OCV**. SOC values normalized by nominal capacity. The capacity fade is observed by a horizontal shift of the curves toward lower SOC values. This capacity fade amounts to approximately 15 %.

**b) ICA**. FOI 1 shifts towards higher voltages (position) and higher ICs (intensity), indicating LLI and possibly LAM$_{PE.}$ FOI 2, the main peak, decreases in intensity from 380 %Q/V to 300 %Q/V, which supports occurring LLI [20,37]. The position of FOI 3 shifts to higher voltages, indicating LLI [37]. The intensity of FOI 4 at higher voltages decreases with LLI [19,20]. Analogously, FOI 5 is decreasing, possibly due to occurring LAM$_{PE}$.

**c) DVA.** The NMC DV consists of a large valley over a wide SOC range, with two characteristic peaks at its sides. These characteristics mainly originate from the graphite anode since the phase transition of the NMC cathode is not reached within the voltage range [40,65]. Therefore, the distance between the two characteristic peaks indicates the storage capability of the anode [40] and is defined as FOI 6. With increasing age, this distance gets slightly smaller, indicating possibly LAM$_{NE}$.

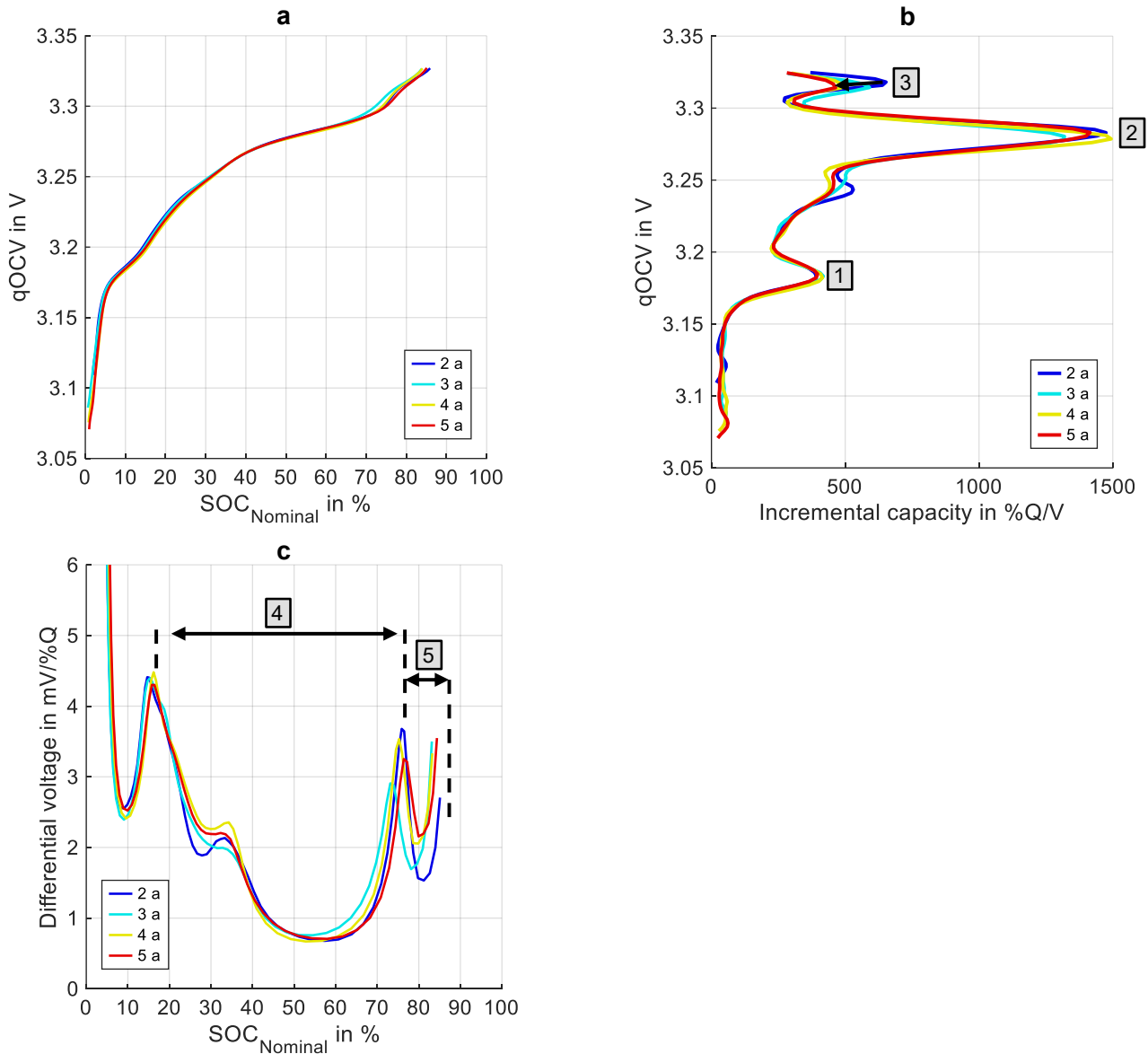

Figure 9: LFP OCV analysis of an exemplary Medium$_{LFP}$ system. The measurements started one year after the HSS installation and ended after four years in 2022.

**a) OCV.** SOC values normalized by nominal capacity. Overall, the LFP curves show very subtle changes over time. With ongoing aging, the curves slightly drop in voltage, following each other closely. Additionally, a small change in curvature can be identified at the beginning and end of the OCV.

**b) ICA.** FOI 1 does not show significant changes, and no conclusion can be drawn. FOI 2 results from the large flat voltages plateau, which can be seen in the OCV curve. It does not show clear trends indicating LLI [37,56]. FOI 3 shows a decline in intensity, which symbolizes LLI [37].

**c) DVA.** The LFP cathode itself does not have clear features in the DV curve due to its flat open circuit potential curve [55]. Therefore, the shape and characteristics seen in the DVA originate mainly from the graphite anode, and the distance between its characteristic peaks allows direct conclusions about its capacity content [66]. The distance of the two peaks represented by FOI 4 stays approximately constant, and does not allow for conclusions [66,67]. FOI 5 possibly indicates LLI as the peak's distance to the steep rise at the EOC voltage slightly decreases.

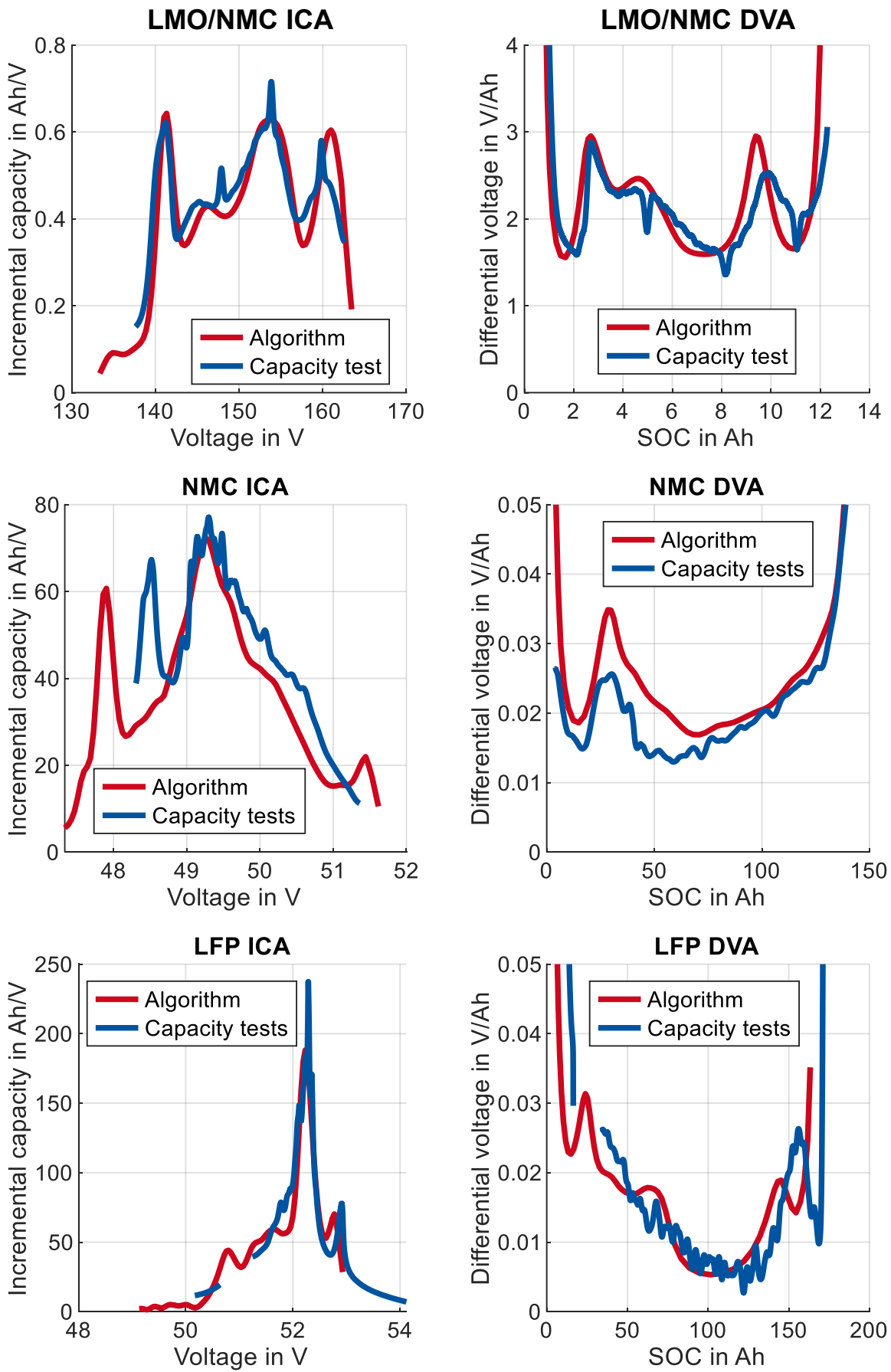

Figure 10: ICA and DVA of exemplary HSSs compared to capacity test.